\documentclass[showpacs,preprintnumbers,amsmath,amssymb, nofootinbib]{revtex4}
\usepackage{fontenc}
\usepackage{graphicx}

\usepackage[dvips]{hyperref}

\usepackage{amsmath}

\begin{document}

\title{Quasinormal modes and Stability Analysis for $4$-dimensional Lifshitz Black Hole}
\author{P.~A.~Gonz\'{a}lez}
\email{pgonzalezm@ucentral.cl} \affiliation{ Escuela de Ingenier\'{i}a Civil en Obras Civiles. Facultad de Ciencias F\'{i}sicas y Matem\'{a}ticas, Universidad Central de Chile, Avenida Santa Isabel 1186, Santiago, Chile}
\affiliation{Universidad Diego Portales, Casilla 298-V, Santiago,
Chile.}
\author{Joel Saavedra}
\email{joel.saavedra@ucv.cl} \affiliation{Instituto de
F\'{\i}sica, Pontificia Universidad Cat\'olica de Valpara\'{\i}so,
Casilla 4950, Valpara\'{\i}so, Chile.}
\author{Yerko V\'{a}squez}
\email{yvasquez@ufro.cl} \affiliation{Departamento de Ciencias F\'{\i}sicas, Facultad de Ingenier\'{\i}a,
Ciencias y Administraci\'{o}n, Universidad de La Frontera, Avenida Francisco
Salazar 01145, Casilla 54-D, Temuco, Chile.}

\date{\today}
\vspace{5cm}

\begin{abstract}

We study the Lifshitz black hole in $4$-dimensions with dynamical exponent $z=2$ and  we calculate analytically the quasinormal modes of scalar perturbations. These quasinormal modes allows to study the stability of the Lifshitz black hole and we have obtained that Lifshitz black hole is stable. 

\end{abstract}

\maketitle

\section{Introduction}

Quasinormal modes (QNMs) of stars and black holes have a long history, from the original works \cite{Regge:1957td, Zerilli:1971wd, Zerilli:1970se} to the modern point of view given in the literature \cite{Kokkotas:1999bd, Nollert:1999ji, Berti:2009kk}. Among the main characteristic of the QNMs, an interesting feature  is their connection to
thermal conformal field theories. According to the AdS/CFT
correspondence~\cite{Maldacena:1997re}, classical gravity
backgrounds in AdS space are dual to conformal field theories at
the boundary (for a review see~\cite{Aharony:1999ti}). Using this
principle it was established that the
relaxation time of a thermal state of the conformal theory at the boundary
is proportional to the inverse of  imaginary part of the QNMs
of the dual gravity background,~\cite{Horowitz:1999jd}. Therefore, the knowledge of the
QNMs spectrum determines how fast a thermal state in the boundary
theory will reach thermal equilibrium. On the other hand, Bekenstein
\cite{Bekenstein:1974jk} was the first who proposed the idea that
in quantum gravity the area of black hole horizon is quantized
leading to a discrete spectrum which is evenly spaced.  About this topic, the QNMs allows to  study  the quantum area
spectrum of the black hole horizon. In those lines an
interesting proposal was made by Hod~\cite{Hod:1998vk} who
conjectured that the asymptotic QNMs frequency is related to the
quantized black hole area. The black hole spectrum can be obtained
imposing the Bohr-Sommerfeld quantization condition to an
adiabatic invariant quantity involving the energy $E$ and the
vibrational frequency $\omega(E)$~\cite{Kunstatter:2002pj}.
Identifying $\omega(E)$ with the real part $\omega_R$ of the QNMs,
the Hod's conjecture leads to an expression of the quantized black
hole area, which however is not universal for all black hole
backgrounds. Furthermore it was argued~\cite{Maggiore:2007nq},
that in the large damping limit the identification of $\omega(E)$
with the imaginary part of the QNMs could lead to the Bekenstein
 universal bound~\cite{Bekenstein:1974jk}.
Therefore, the analytic computation of QNMs for black holes geometry is an important task in order to connect different key ideas about black hole physics and gravitational theories. Recently another consequence of  AdS/CFT
correspondence~\cite{Maldacena:1997re} has been the application beyond high energy physics to another areas of physics, for example quantum chromodynamics and condensed matter. In this sense, Lifshitz spacetimes have received great attention from the condensed matter point of view i.e., the searching of gravity duals of Lifshitz fixed points. From quantum field theory point of view, there are many  scale invariant theories which are of interest in studying such critical points. These kind of theories exhibit the anisotropic scale invariance $t\rightarrow \lambda ^z t$, $x\rightarrow \lambda x$, with $z\ne1$, where $z$ is the relative scale dimension of time and space and they are of particular interest in the studies of critical exponent theory and phase transition.  Lifshitz spacetimes, described by the metric
\begin{equation}
ds^2=- l^2 \left(  r^{-2z}dt^2+r^{-2}dr^2+r^{-2}d\vec{x}^2 \right), \label{lif1}
\end{equation} 
where $\vec{x}$ represents a $D-2$ spatial vector and $l$ denotes the  length scale in the geometry. 

In this work we will consider  a matter distribution outside the horizon of one Lifshitz black hole, we will focus our study in the analytic solution found in Ref.~\cite{Balasubramanian:2009rx}. The action is 
\begin{equation}
S=\frac{1}{2}\int {d^4x(R-2\Lambda)}-\int {d^4x\left( \frac{e^{-2\phi}}{4}F^2+\frac{m^2_A}{2}A^2+(e^{-2\phi}-1)\right)}~,
\end{equation}
where, $R$ is the Ricci scalar, $\Lambda$ is the cosmological constant, $A$ is a gauge field, $F$ is the field strength and  $m^2_A=2z$. The black hole solution of this system that asymptotically approaches to the Lifshitz spacetime is given by
\begin{eqnarray}
\nonumber\phi=-\frac{1}{2}log\left(1+\frac{r^2}{r_H^2}\right)~,\\
\nonumber A=\frac{f ( r )}{r^2}dt~,\\
ds^2=-f ( r )\frac{dt^2}{r^{2z}}+\frac{d\vec{x}^2}{r^2}+\frac{dr^2}{f ( r )r^2}~,
\end{eqnarray}
with
\begin{equation}
f ( r )=1-\frac{r^2}{r_H^2}~.
\end{equation}
For $z=2$, the metric of Lifshitz black hole is 
\begin{equation}\label{metric2}\
ds^{2}=-\frac{f( r )}{r^4}dt^{2}+\frac{1}{f( r )r^2}dr^{2}+\frac{1}{r^{2}}d\vec{x}^{2}~.
\end{equation}

In order to compute the QNMs, we  perturbed this  geometry with a scalar field and also we are assuming that the back reaction is vanished over this geometry. Then we just need to solve the Klein-Gordon equation of scalar field for the geometry under consideration. Therefore, we obtain the analytic solution of the quasinormal modes which are characterized by a spectrum that is independent of the initial conditions of the perturbation and depends only on the black hole parameters and on the fundamental constants of the system (for a recent review see~\cite{Siopsis:2008xz,Konoplya:2011qq}) and also through our analytic results we study the stability of the Lifshitz black hole via QNMs.

The paper is organized as follows. In Sec. II we calculate the
exact QNMs of the scalar perturbations of 4-dimensional
Lifshitz black hole. Our conclusions are in Sec. III.



\section{Quasinormal Modes}
The quasinormal modes of scalar
perturbations  for a minimally coupled scalar field to curvature on the background of a Lifshitz black hole in  four dimensions are described by the solution of Klein-Gordon equation
\begin{equation}
\Box \psi = \frac{1}{\sqrt{-g}}\partial_{\mu}\left(\sqrt{-g} g^{\mu\nu}\partial_{\nu}\right)\psi=m^2\psi~, \label{k-g}
\end{equation}
where $m$ is the mass of the scalar field $\psi$. In order to solve Klein-Gordon equation we adopt the ansatz $\psi = R( r )e^{-i\vec{\kappa} \cdot \vec{x}}e^{-i\omega t}$. Thus, Eq. (\ref{k-g}) gives
\begin{equation}\label{radialr}\
\partial_{r}^2R( r )-\frac{1}{r}\left(1+\frac{2}{1-\frac{r^2}{r^2_H}}\right)\partial_{r}R(r)+
\frac{1}{1-\frac{r^2}{r^2_H}}\left(-\kappa^2+\frac{r^2\omega^2}{1-\frac{r^2}{r^2_H}}-\frac{m^2}{r^2}\right)R( r )=0~.
\end{equation}

\begin{center}
\begin{figure}
\includegraphics[width=4.0in,angle=0,clip=true]{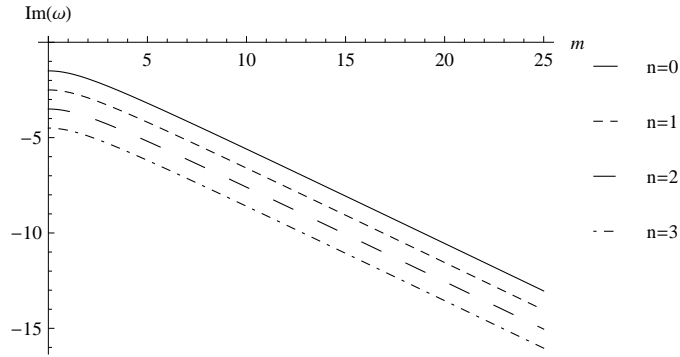}
\caption{Im( $\omega$) v/s $m$; $r_H=1$ and $\kappa=1$.}
\label{modos7}
\end{figure}
\end{center}

\begin{center}
\begin{figure}
\includegraphics[width=4.0in,angle=0,clip=true]{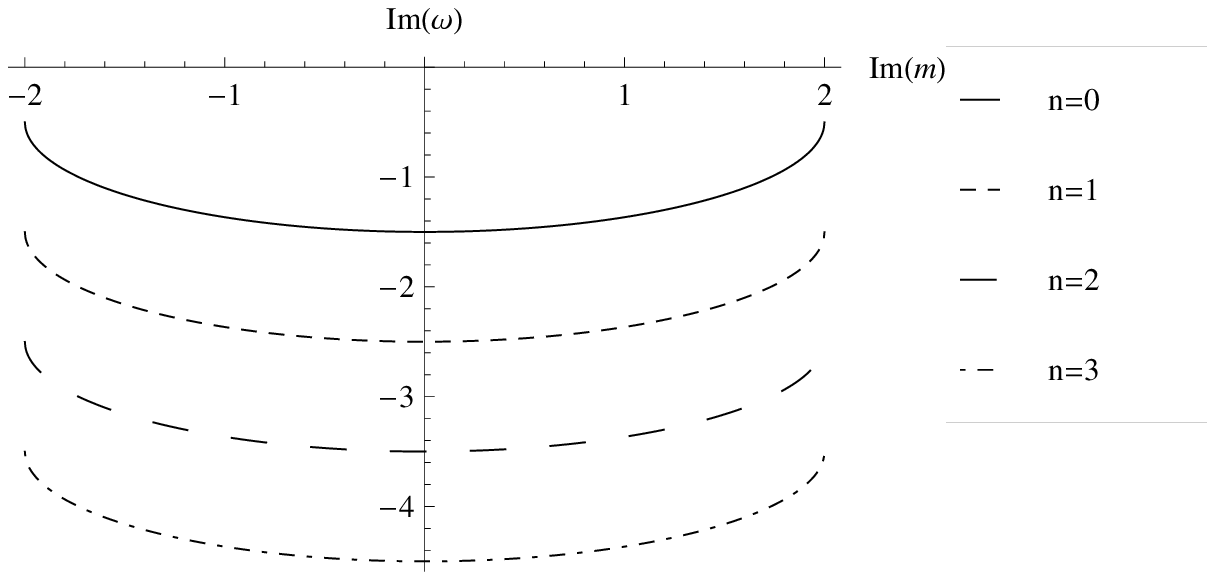}
\caption{Im( $\omega$) v/s $Im( $m$)$; $r_H=1$ and $\kappa=1$.}
\label{modos77}
\end{figure}
\end{center}

Performing the change of variables $u=\frac{r}{r_H}$ and inserting in Eq. (\ref{radialr})  we obtain
\begin{equation}\label{radialu}
\partial_{u}^2R(u)+\frac{u^2-3}{u(1-u^2)}\partial_{u}R(u)+
\frac{r_H^2}{1-u^2}\left(-\kappa^2+\frac{\omega^2u^2r_H^2}{1-u^2}-\frac{m^2}{u^2r_H^2}\right)R(u)=0~,
\end{equation}
now using the changes $v=u^2$ and $1-v=z$, the Eq.~(\ref{radialu}) can be written as
\begin{equation}
\partial_{z}^2R(z)+\left[\frac{1}{1-z}+\frac{1}{z}\right]\partial_{z}R(z)+
\frac{r_H^2}{4z(1-z)}\left(-\kappa^2+\frac{\omega^2r_H^2(1-z)}{z}-\frac{m^2}{r_H^2(1-z)}\right)R(z)=0~,
\end{equation}
which, under the decomposition $R(z)=z^\alpha(1-z)^\beta K(z)$ with
\begin{equation}
\alpha_{\pm}= \pm \frac{ir_H^2\omega}{2}~,
\end{equation}
\begin{equation}
\beta_{\pm}= \frac{1}{2}\left(2\pm \sqrt{4+m^2}\right)~,
\end{equation}
can be written as a hypergeometric equation for K
\begin{equation}\label{hypergeometric2}
z(z-1)K''(z)+\left[c-(1+a+b)z\right]K'(z)-abK(z)=0~.
\end{equation}
Where 
\begin{equation}
c=1+2\alpha~,
\end{equation}
\begin{equation}
a=\frac{1}{2}\left(-1 \pm \sqrt{1-r_H^2\kappa^2-r_H^4\omega^2}+2\alpha+2\beta\right)~,
\end{equation}
\begin{equation}
b=\frac{1}{2}\left(-1 \mp \sqrt{1-r_H^2\kappa^2-r_H^4\omega^2}+2\alpha+2\beta\right)~.
\end{equation}
The general solution of Eq. (\ref{hypergeometric2}) takes
the form
\begin{equation}
K=C_{1}F_{1}(a,b,c;z)+C_2z^{1-c}F_{1}(a-c+1,b-c+1,2-c;z)~,
\end{equation}
which has three regular singular point at $z=0$, $z=1$ and
$z=\infty$. Here, $F_{1}(a,b,c;z)$ is a hypergeometric function
and $C_{1}$, $C_{2}$ are constants. Then, the analytic solution for the
radial function $R(z)$ is
\begin{equation}\label{RV}\
R(z)=C_{1}z^\alpha(1-z)^\beta
F_{1}(a,b,c;z)+C_2z^{-\alpha}(1-z)^\beta
F_{1}(a-c+1,b-c+1,2-c;z)~.
\end{equation}
So, in the vicinity of the horizon, $z=0$ and using
the property $F(a,b,c,0)=1$, the radial function $R(z)$ behaves as
\begin{equation}
R(v)=C_1 e^{\alpha \ln z}+C_2 e^{-\alpha \ln z},
\end{equation}
and the scalar field $\psi$ can be written as
\begin{equation}
\psi\sim C_1 e^{-i\omega(t+ \frac{r_H^2}{2}\ln z)}+C_2
e^{-i\omega(t-\frac{r_H^2}{2}\ln z)}~, \label{psi1}
\end{equation}
in which, without loss of generality we considers  $\alpha=\alpha_-$, the first term in Eq. (\ref{psi1}) represents an ingoing wave and the second one an outgoing wave in the black hole geometry. In order to compute the QNMs, we
must to impose  boundary conditions on the horizon, 
the existence only the ingoing waves (this fixes  $C_2=0$). Then, radial
solution becomes
 \begin{equation}\label{horizonsolution}
R(z)=C_1 e^{\alpha \ln z}(1-z)^\beta F_{1}(a,b,c;z)=C_1
e^{-i\omega\frac{r_H^2}{2}\ln z}(1-z)^\beta F_{1}(a,b,c;z)~.
\end{equation}
Now, we need to implement boundary conditions at infinity ($z=1$), before to perform this task we
shall apply to Eq. (\ref{horizonsolution}) the Kummer's formula
for the hypergeometric function as a connection formula \cite{M. Abramowitz},
\begin{equation}\label{relation}
F_{1}(a,b,c;z)=\frac{\Gamma(c)\Gamma(c-a-b)}{\Gamma(c-a)\Gamma(c-b)}
F_1(a,b,a+b-c,1-z)+(1-z)^{c-a-b}\frac{\Gamma(c)\Gamma(a+b-c)}{\Gamma(a)\Gamma(b)}F_1(c-a,c-b,c-a-b+1,1-z),
\end{equation}
with this expression the radial function reads
\begin{eqnarray}\label{R}\
\nonumber R(v) &=& C_1 e^{-i\omega\frac{r_H^2}{2}\ln z}(1-z)^\beta\frac{\Gamma(c)\Gamma(c-a-b)}{\Gamma(c-a)\Gamma(c-b)} F_1(a,b,a+b-c,1-z)\\
&& +C_1 e^{-i\omega\frac{r_H^2}{2}\ln
z}(1-z)^{c-a-b+\beta}\frac{\Gamma(c)\Gamma(a+b-c)}{\Gamma(a)\Gamma(b)}F_1(c-a,c-b,c-a-b+1,1-z)~.
\end{eqnarray}
For,  $\beta_+>2$ the field at infinity is regular if  the gamma function $\Gamma(x)$ has the poles at $x=-n$ for
$n=0,1,2,...$, the wave function satisfies the considered boundary
condition only upon the following additional restriction
$(a)|_{\beta_+}=-n$ or $(b)|_{\beta_+}=-n$. In the cases of  $\beta_-<0$ the field at infinity is regular or null if  the gamma function $\Gamma(x)$ has the poles at $x=-n$ for
$n=0,1,2,...$, the wave function satisfies the considered boundary
condition only upon the following additional restriction
$(c-a)|_{\beta_-}=-n$ or $(c-b)|_{\beta_-}=-n$. These conditions
determine the form of the frequency of quasinormal modes as 
\begin{equation}
\omega=-i\left[\frac{r_H^2 k^2 (-1 - 2 n) + m^2 (1+ 2 n) + 
 2 (2 + 2 n - 6 n^2 - 4 n^3)}{2r_ H^2 (3 + m^2 - 4 n - 4 n^2) }+\frac{\sqrt{4 + m^2}(r_H^2\kappa^2+m^2+2-4n-4n^2)}{2r_ H^2 (3 + m^2 - 4 n - 4 n^2)}\right]~.
\end{equation}
The stability of black holes can be analyzed via QNMs and it is known that such stability is guaranteed if the imaginary part of the QNMs is negative. Now, we consider the QNMs  that we have analytic found in the previous section. In this way and without loss of generality we have chosen $r_H=1$ and $\kappa =1$ and we have plotted in Fig.~(\ref{modos7}), the imaginary part of the modes with respect to the scalar fields mass. 
Till this point we have considered a positive value for the scalar field mass. Now, we are considering a scalar field with imaginary mass and we plot the behavior of the quasinormal modes with respect to imaginary part of the scalar field mass in  Fig.~(\ref{modos77}). 
Note that from Fig.~(\ref{modos77}), for $(-2<m^2<2)$,  the stability of the Lifshitz black hole is guaranteed .  
It is worth noting, that the condition on the  $m^2$ that we have used agrees
with the analogue to Breitenlohner-Freedman condition,   that any effective mass
must be satisfied in order to have a stable propagation Ref.~\cite{Kachru:2008yh}.

\section{Conclusions}

Lifshitz black hole is a very interesting static solution of gravity theories which asymptotically approach to Lifshitz spacetimes. This spacetime is of interest due to its  invariance under anisotropic scale transformation and its representation as the gravitational dual of strange metals \cite{Hartnoll:2009ns}. If $z=1$, the spacetime is the usual AdS metric in Poincare coordinates. Furthermore, all scalar curvature invariants are constant and there is a curvature singularity for $z\ne 1$, Ref.~\cite{{Hartnoll:2009sz},{Kachru:2008yh},{Copsey:2010ya},{Horowitz:2011gh}}. 
In this work we calculated the analytic QNMs of scalar perturbations for the four dimensional Lifshitz black hole with plane topology by using Dirichlet boundary condition at infinity. First, we have found that the quasinormal modes are purely imaginary i.e., there is only damped modes. Second, using our analytic results we studied the stability of this geometry under scalar perturbation, we note that the stability is guaranteed if the imaginary part of the QNMs is negative. In this sense, our results are summarized  in Figs.~(\ref{modos7}) and (\ref{modos77}), where the imaginary part of the modes is plotted with respect to the scalar fields mass. For the cases of positive mass we found, from modes with $n \geq 0$ the imaginary part of QNMs frequency satisfy the stability condition. Also, we considered a scalar field with imaginary mass and we found that for $(-2<m^2<2)$,  the stability of the Lifshitz black hole is guaranteed .  
We allow to note, that the condition on the  $m^2$ that we have used agrees
with the analogue to Breitenlohner-Freedman condition.


\section*{Acknowledgments}
We thank Bertha Cuadros-Melgar for enlightening discussions and remarks.
Y.V. was supported by Direcci\'{o}n de Investigaci\'{o}n y Desarrollo,
Universidad de la Frontera, DIUFRO DI11-0071. J.S. was supported by  COMISION NACIONAL DE CIENCIAS Y
TECNOLOGIA through FONDECYT Grant 1110076, 1090613 and 1110230. This work was also
partially supported by PUCV DI-123.713/2011 (JS). P.G. acknowledge the hospitality of the Physics Department of Universidad de La Frontera where part of this work was made.


\appendix

\end{document}